\documentclass[twocolumn]{revtex4-1}

\usepackage{graphicx,amsmath}

\begin{document}

\title{Topological edge states in equidistant arrays of Lithium Niobate nano-waveguides}

\author{Andrey V. Gorbach}
\affiliation{Centre for Photonics and Photonic Materials, Department of Physics, University of Bath, Bath BA2 7AY, United Kingdom}
\email{Corresponding author: A.Gorbach@bath.ac.uk}
\author{Jesper Beer}
\affiliation{Department of Physics, University of Bath, Bath BA2 7AY, United Kingdom}
\author{Anton Souslov}
\affiliation{Department of Physics, University of Bath, Bath BA2 7AY, United Kingdom}

\begin{abstract}
We report that equidistant 1D arrays of thin-film Lithium Niobate nano-waveguides generically support topological edge states. Unlike conventional coupled-waveguide topological systems, the topological properties of these arrays are dictated by the interplay between intra- and inter-modal couplings of two families of guided modes with different parities. Exploiting two modes within the same waveguide to design a topological invariant allows us to decrease the system size by a factor of two and substantially simplify the structure. We present two example geometries where topological edge states of different types (based on either quasi-TE or quasi-TM modes) can be observed within a wide range of wavelengths and array spacings.
\end{abstract}

\maketitle

Topological photonic systems have recently attracted much attention, not only as a potential playground to explore fundamental physical effects associated with topological states, but also as a new platform to design structures for light manipulation~\cite{Lu2014, Ozawa2019, Lan2022}. 
Of particular interest are recently emerging all-dielectric structures, whereby topological properties are defined by the structure of a photonic crystal~\cite{Wu2015a, Wang2022a}.
Remarkably, non-trivial topological phases may exist even in simple 1D crystals~\cite{Lang2012}. A fundamental workhorse of topological physics is a 1D system called the Su-Schrieffer-Heeger (SSH) chain~\cite{Su1980}. This model, which was originally proposed to describe excitations in polyacetylene molecules, represents a 1D dimer chain with alternating coupling (hopping) coefficients. The two topologically distinct phases of the chain correspond to two different configurations where either the stronger or the weaker coupling defines the unit cell~\cite{Li2014a}. The topological invariant that distinguishes these phases is known as either the winding number or the Zak phase~\cite{Zak1989}. One important manifestation of topological phases in 1D systems is the emergence of edge states~\cite{Lang2012}. Existence of such localized states is directly related to the topological properties of the bulk crystal through bulk-boundary correspondence~\cite{Hatsugai1993,Xiao2014}. This correspondence dictates that, because the invariant has to abruptly change at the boundary of the topological material, this boundary is required to host protected edge states.

The vast majority of photonic topological structures explored so far impose the required crystal symmetry by spatially modulating the dielectric constant. This approach stems from a general analogy between condensed matter physics and photonics~\cite{Photonic_Crystals_textbook}. Particularly, the standard models describing light propagation in coupled waveguide systems directly map onto tight-binding models, such as the SSH chain~\cite{Blanco-Redondo2016}. Such models assume single-mode operation of the waveguides. Recently, an alternative approach has been proposed, whereby the multi-modeness of the coupled waveguides is exploited to expand the number of degrees of freedom per unit cell~\cite{Savelev2020, Bobylev2021, Caceres-Aravena2020}. Here, the effective crystal structure and its topological properties are governed by the network of different intra- and inter-modal couplings, while the spatial arrangement of the waveguides can be entirely homogeneous. Thus, the complexity of topological photonic bands can be realised in much simpler and more compact structures.

\begin{figure}
    \centering
    \includegraphics[width=\linewidth]{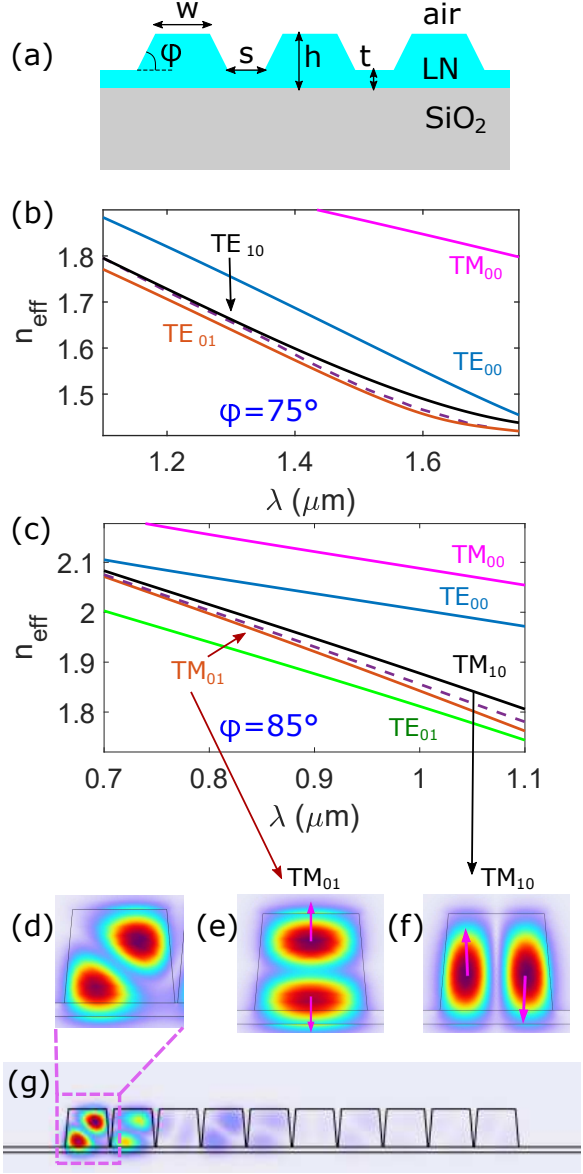}
    \caption{Edge states in LNOI waveguide arrays: (a) a schematic view of an array; (b) dispersion of different guided modes in a single waveguide with $w=700\textrm{nm}$, $h=800\textrm{nm}$, $t=100\textrm{nm}$, $\varphi=75^\circ$, using x-cut LN film (the extraordinary axis of the crystal is oriented horizontally). The dashed line illustrates dispersion of the edge mode in an array of $N=10$ waveguides, edge-to-edge separations $s=100\textrm{nm}$; (c) the same as (b) but for $\varphi=85^\circ$, dashed line illustrates dispersion of the edge mode with $s=50nm$; (d)-(f) field profiles (norm of the electric field) of guided modes in the same geometry as in panel (c), with $\lambda=0.7\mu\textrm{m}$. An edge mode in waveguides array with $N=10$ and $s=50\textrm{nm}$  is shown in panel (g), and a zoom-in of the waveguide at the edge is displayed in (d). Parts (e) and (f) show profiles of quasi-$\textrm{TM}_{01}$ and quasi-$\textrm{TM}_{10}$ modes of an isolated waveguide, respectively. The arrows indicate the polarization of the local electric field. (Data from COMSOL Multiphysics).}
    \label{fig:fig1}
\end{figure}

In this work, we demonstrate that one-dimensional equidistant (homogeneous) arrays of Lithium Niobate on Insulator (LNOI) waveguides~\cite{Poberaj2012, Boes2018, Zhu2021} can exhibit topologically distinct phases, leading to formation of topological edge states, see Fig.~\ref{fig:fig1}. Ridge waveguides are etched from a Lithium Niobate (LiNbO$_3$) film of thickness $h$ on a silica glass substrate. The waveguides are characterised by width $w$ at the top, residual film thickness $t$, and sidewall angle $\varphi$ [Fig.~\ref{fig:fig1}(a)]. This angle typically varies from  $40^\circ$ to $80^\circ$, depending on the particular etching process~\cite{Zhu2021}. Combined with the anisotropic dispersion of bulk Lithium Niobate, the four geometrical parameters ($h,t,w,\varphi$) represent a convenient toolbox for tuning the dispersion of an isolated waveguide. Particularly, for certain parameters one can observe a nearly degenerate behaviour of different pairs of guided modes within large spectral windows. One such example geometry is illustrated in Fig.~\ref{fig:fig1}(b), where two such modes, labelled quasi-$\textrm{TE}_{01}$ and quasi-$\textrm{TE}_{10}$, have similar effective indices within a wide wavelength interval $1.0\mu\textrm{m}<\lambda<1.7\mu\textrm{m}$ (these modes would be completely degenerate in a perfect square waveguide). Adjusting the sidewall angle, a similar nearly-degenerate behaviour of quasi-$\textrm{TM}_{01}$ and quasi-$\textrm{TM}_{10}$ modes can be observed, see Fig.~\ref{fig:fig1}(c). This trend appears to be generic: similar nearly-degenerate behaviour of different pairs of modes can be observed in LNOI waveguides by varying the three geometrical parameters. The presence of two degenerate modes within each waveguide is the key ingredient that enables topological states within an equidistant array.

\begin{figure}
    \centering
    \includegraphics[width=\linewidth]{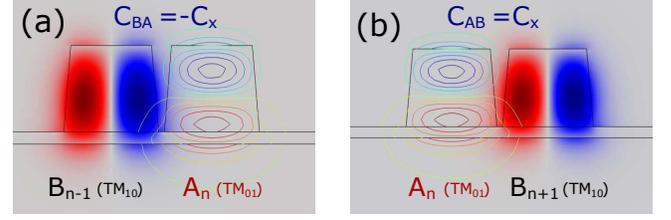}
    \caption{Overlaps between quasi-$\textrm{TM}_{01}$ mode in waveguide $n$ and quasi-$\textrm{TM}_{10}$ mode in waveguides $n-1$ (a) and $n+1$ (b) entering the calculation of the coupling constant in Eq.~(\ref{eq:overlap}).}
    \label{fig:fig2}
\end{figure}

Arranging such waveguides in a regular 1D array, as in Fig.~\ref{fig:fig1}(a), we observe the formation of  localized edge modes within a wide range of wavelengths and edge-to-edge separation distances $s$ between the waveguides. Figure~\ref{fig:fig1}(g) shows one example of an edge mode in an array of waveguides with the same parameters as in Fig.~\ref{fig:fig1}(c). In this mode, the field intensity is exponentially localized within a few waveguides nearest to the edge of the array. This mode is doubly-degenerate: an equivalent "mirror" mode exists on the opposite edge. Similar edge modes are observed in the second geometry. Dashed lines in Fig.~\ref{fig:fig1}(b) and (c) show dispersions of the edge modes in finite-size arrays composed of waveguides having the two respective geometries. In both cases and for all wavelengths, the effective index of an edge mode appears to be in-between the indices of the two nearly degenerate modes of a single waveguide.

A close inspection of the field distribution of the edge mode in Fig.~\ref{fig:fig1}(g) within the area of the first waveguide, see Fig.~\ref{fig:fig1}(d), reveals that a superposition of the quasi-$\textrm{TM}_{01}$ and quasi-$\textrm{TM}_{10}$ modes is excited within this waveguide.  These numerical results were obtained from COMSOL Multiphysics taking into account the material dispersion of both the Lithium Niobate and the silica glass substrate~\cite{Cai2018}. The corresponding mode profiles of an isolated waveguide are shown in Figs.~\ref{fig:fig1}(e) and~\ref{fig:fig1}(f). Thus, substracting the fields of quasi-$\textrm{TM}_{01}$ and quasi-$\textrm{TM}_{10}$ modes results in the diagonal structure observed in Fig.~\ref{fig:fig1}(d). A similar structure is observed in other waveguides, see Fig.~\ref{fig:fig1}(g), and in edge modes supported by the second geometry in Fig.~\ref{fig:fig1}(b).

\begin{figure*}
    \centering
    \includegraphics[width=\textwidth]{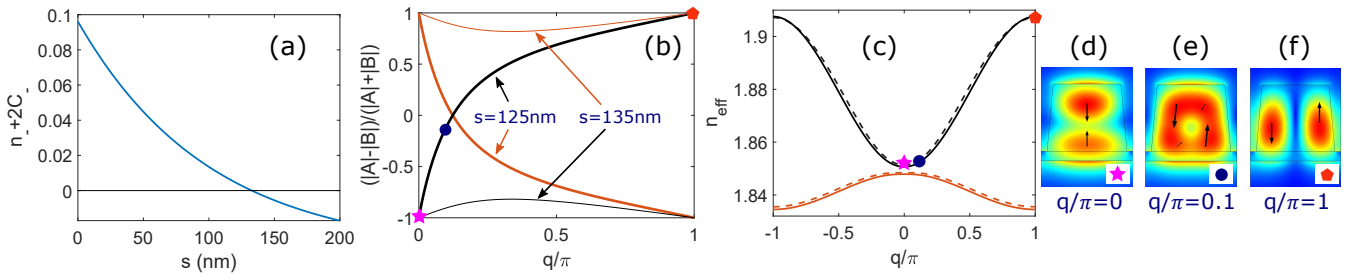}
    \caption{Band structure of an infinite array with the same waveguide parameters as Fig.~\ref{fig:fig1}(b): (a) the balance between the modal detuning and coupling coefficients, $n_-+2C_-$, as a function of the waveguide separation at $\lambda=1.0\mu\text{m}$. The gap closes at $s\approx 130$nm; (b) the structure of eigenvectors of the coupled-mode model in Eqs.~(\ref{eq:An},\ref{eq:Bn}) of the top (black) and bottom (red) bands for $s=125$nm (thick lines) and $s=135$nm (thin lines); (c) the band structure for $s=125$nm obtained using Comsol simulations (solid curves) and the coupled-mode model (dashed curves); (d)-(f) profiles of the modes of the top band for $s=125$nm at $q/\pi=0$, $0.1$, and $1$, respectively.}
    \label{fig:fig3}
\end{figure*}

What is the origin of these edge states? To answer this question, we consider a simple coupled-mode model, which takes into account interactions between the two different types of modes for each waveguides:
\begin{eqnarray}
\nonumber
-i\frac{d A_n}{dz}&=&n_A A_n + C_A\left(A_{n+1}+A_{n-1}\right)\\
\label{eq:An}
&&+C_x\left(B_{n+1}-B_{n-1}\right)
\;,\\
\nonumber
-i\frac{d B_n}{dz}&=&n_B B_n + C_B\left(B_{n+1}+B_{n-1}\right)\\
\label{eq:Bn}
&&-C_x\left(A_{n+1}-A_{n-1}\right)
\;.
\end{eqnarray}
Here $A_n$ and $B_n$ are amplitudes of the two modes [e.g., quasi-$\textrm{TE}_{01}$ and quasi-$\textrm{TE}_{10}$ for the geometry in Fig.~\ref{fig:fig1}(b)] in the $n$-th waveguide, $z$ is the dimensionless propagation length measured in the units of the wavelength in vacuum $\lambda_0=2\pi c/\omega$, $n_A$ and $n_B$ are the effective indices of the two modes for an isolated waveguide, and $C_A$, $C_B$, and $C_x$ are different intra- and inter-modal coupling coefficients. One important aspect of this model is the variation of signs of the inter-modal coupling coefficient $C_x$ connecting mode $A_n$ and mode $B_{n+1}$ ($C_x$), and connecting mode $A_n$ and mode $B_{n-1}$ ($-C_x$). 
This is
a direct consequence of the opposite parities of the two interacting modes~\cite{Caceres-Aravena2020}. Generally, the coupling coefficient between mode $p$ in the waveguide $n$ and mode $q$ in the waveguide $(n+1)$, with $p$ and $q$ each being either mode $A$ or mode $B$, is obtained via the overlap integral~\cite{liu_2016}:
\begin{equation}
    \label{eq:overlap}
    C_{pq}=\omega \iint_{-\infty}^{+\infty} \vec{e}^*_p(x,y)\cdot\Delta\epsilon\vec{e}_q(x+T,y)dxdy\;,
\end{equation}
where $\vec{e}_{A,B}(x,y)$ are the modes of the isolated waveguide $n$, $T$ is the centre-to-centre distance between the waveguides, and $\Delta\epsilon(x,y)$ is the difference between the permittivity tensor of the two-waveguide structure (waveguides $n$ and $n+1$) and a single-waveguide structure (waveguide $n$ only). Essentially, $\Delta\epsilon$ is non-zero within the core area of the $(n+1)$th waveguide only. For the modes of the same type, i.e., when $p=q$, the coupling coefficients between pairs of waveguides $n$ and $(n+1)$, and between waveguides $n$ and $(n-1)$, will be the same. However, this is no longer the case if the modes are of different types. In particular, when the two modes have opposite symmetries with respect to 
$x\to -x$, such as quasi-$\textrm{TM}_{01}$ and quasi-$\textrm{TM}_{10}$ modes, one obtains $C_{pq}=-C_{qp}$, as illustrated in Fig.~\ref{fig:fig2}. 
Notably, this variation of signs preserves the Hermitian structure of the model, but induces an effective chirality in the array.

\begin{figure*}
    \centering
    \includegraphics[width=\textwidth]{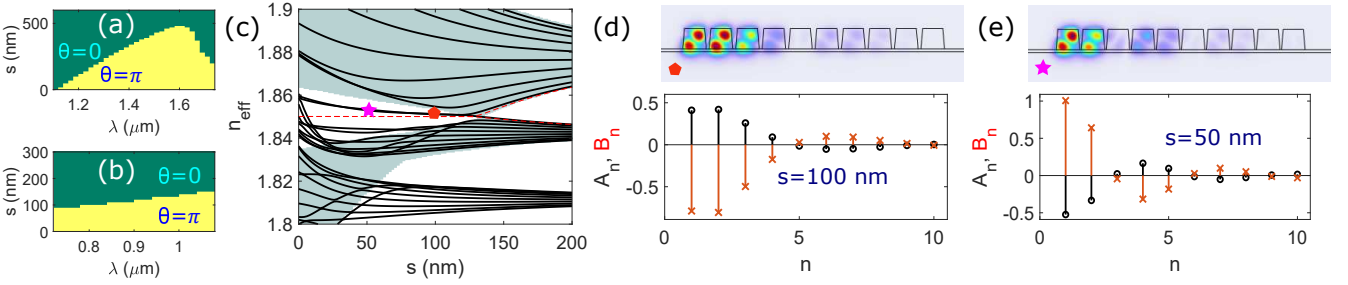}
    \caption{(a) and (b) Zak phase of the top band  (corresponding to $\textrm{TE}_{10}$/$\textrm{TM}_{10}$ modes for sufficiently large $s$) of an infinite waveguide array,  Eq.~(\ref{eq:zak}), evaluated for the geometries as in Fig.~\ref{fig:fig1}(b) and (c), respectively. The light shaded areas indicate the regions with $\theta=\pi$, where we predict the existence of edge modes; (c) modes of a finite waveguides array with $N=10$ for the geometry as in Fig.~\ref{fig:fig1}(c) and with  $\lambda=1\mu\text{m}$, obtained from COMSOL simulations. The red dashed lines show two modes of the coupled-modes model, Eqs.~(\ref{eq:An},\ref{eq:Bn}), which correspond to edge states. The shaded areas indicate bandwidths of the two bands of the coupled-modes infinite system, Eq.~(\ref{eq:spectr_discr}); (d) and (e) profiles of edge modes in finite waveguide arrays with $N=10$, $s=100$nm and $s=50$nm. The top panels are the results of COMSOL simulations, the corresponding effective indices are marked with the star and pentagon symbols in panel (c). The bottom panels are the eigenmodes of the model in Eqs.~(\ref{eq:An},\ref{eq:Bn}).}
    \label{fig:fig4}
\end{figure*}

For an infinite array, the spectrum of the model in Eqs.~(\ref{eq:An},\ref{eq:Bn}) for plane waves $A_n,B_n\sim \exp(i\lambda z - iqn)$ consists of two bands: 
\begin{eqnarray}
    \nonumber
    \lambda_{1,2}&=&n_+ +2C_+\cos(q)\\
    \label{eq:spectr_discr}
    &&\pm\sqrt{(n_-+2C_-\cos(q))^2+4C_x^2 \sin^2 q}\;,
\end{eqnarray}
where 
$n_\pm=(n_B\pm n_A)/2$ and $C_\pm=(C_B\pm C_A)/2$. Notably, the gap between the bands closes at $q=0$ when $n_-+2C_-=0$, or at $q=\pi$ when $n_--2C_-=0$. This gap closure is accompanied by a qualitative change in the structure of the eigenvectors, as illustrated in Figs.~\ref{fig:fig3}(a) and (b). Here, for the geometry as in Fig.~\ref{fig:fig1}(c), by varying the separation distance between the wavegudes at a fixed wavelength, the balance between $|n_-|$ and $2|C_-|$ is tipped over. When $|n_-|>2|C_-|$ ($s>s_0\approx 130$nm),
the amplitudes of either $A$ or $B$ modes dominate across the entire Brillouin zone $0\le|q|\le\pi$ in each band, see the thin lines in Figs.~\ref{fig:fig3}(b). Thus, each band can be associated with a particular mode ($A$ or $B$) in this case. On the contrary, when $|n_-|<2|C_-|$ ($s<s_0$), the structure of the modes within each band switches between mode $A$ and $B$ as the wavenumber $q$ sweeps the Brillouin zone, see the thick lines in Fig.~\ref{fig:fig3}(b). These results of the coupled-mode model are in agreement with the full solution of Maxwell's equations with periodic boundary conditions. In Fig.~\ref{fig:fig3}(c), a part of the spectrum of an infinite waveguide array (in the vicinity of quasi-$\textrm{TM}_{10}$ and quasi-$\textrm{TM}_{01}$ modes for an isolated waveguide) is shown as obtained using COMSOL Multiphysics (solid curves). The corresponding spectrum of the coupled-mode model is shown with dashed curves. For the latter, we used COMSOL data for isolated waveguides to calculate the coupling coefficients according to Eq.~(\ref{eq:overlap}). The profiles of the modes of the top band at different wavenumbers $q$ are shown in Figs.~\ref{fig:fig3}(d)-(f). As predicted by the coupled-modes model, we observe a transition from quasi-$\textrm{TM}_{01}$ at $q=0$ to quasi-$\textrm{TM}_{10}$ at $q=\pi$.

For the model in Eqs.~(\ref{eq:An},\ref{eq:Bn}), it was demonstrated that, as the spectral gap closes and reopens again, the system undergoes a topological transition~\cite{Caceres-Aravena2020}. The same is true for the full model, as we confirm by calculating the Zak phase of the two bands using the Wilson loop approach~\cite{Wang2019}:
\begin{equation}
\label{eq:zak}
     \theta\approx i\ln\Pi_{i=1}^{N}\left<\psi(k_i),\psi(k_{i+1})\right>\;,
\end{equation}
where $\psi(k)$ are the normalized eigen-modes belonging to a particular band, and the inner product of two modes is defined as
\begin{equation}
     \left<a,b\right>=
     \frac14\iint \left[\vec{e}_a\times \vec{h}^*_b+\vec{e}^*_b\times \vec{h}_a\right]dxdy\;.
\end{equation}
The evaluation in Eq.~(\ref{eq:zak}) is performed by discretizing the full Brillouin zone into $N$ segments with $k_{N+1}=k_1=-\pi$. As the separation between the waveguides crosses the transition point $s=s_0$, we observe a jump from $\theta=0$ (trivial phase corresponding to winding number $0$) to $\theta=\pi$ (non-trivial phase corresponding to winding number $1$) in each of the two bands. In Figs.~\ref{fig:fig4}(a) and (b), the Zak phase of the top band is plotted for the same geometries as in Figs.~\ref{fig:fig1}(b) and (c), respectively. This topological phase transition is accompanied by the emergence of two degenerate edge modes (localized at either edge) in a finite-size array, as illustrated in Figs.~\ref{fig:fig4}(c)--(e). In Fig.~\ref{fig:fig4}(c), the spectrum of an infinite size coupled-modes model, Eq.~(\ref{eq:spectr_discr}), is shown with the shaded areas for the same geometry as in Fig.~\ref{fig:fig1}(c) at $\lambda=1\mu\textrm{m}$. The modes of a finite-size array ($N=10$ waveguides) are shown with solid (Comsol simulations) and dashed (coupled-modes model) lines. As the gap of an infinite system closes and re-opens, the two modes corresponding to the bottom and top edges of the two bands at $s>s_0$ merge together to form the two degenerate edge states at $s<s_0$. In the coupled-modes model, these two states have a fixed effective index for any $s$, see the dashed lines in Fig.~\ref{fig:fig4}(c). In the full system, the indices are no longer fixed due to the influence of a third band corresponding to quasi-$\textrm{TE}_{01}$ modes, c.f.~Fig.~\ref{fig:fig1}(c). Nevertheless, the coupled-modes model gives a reasonably accurate prediction, not only for the effective index, but also for the detailed field profiles of the edge modes, as shown in Figs.~\ref{fig:fig4}(d) and (e).

\begin{figure*}
    \includegraphics[width=\textwidth]{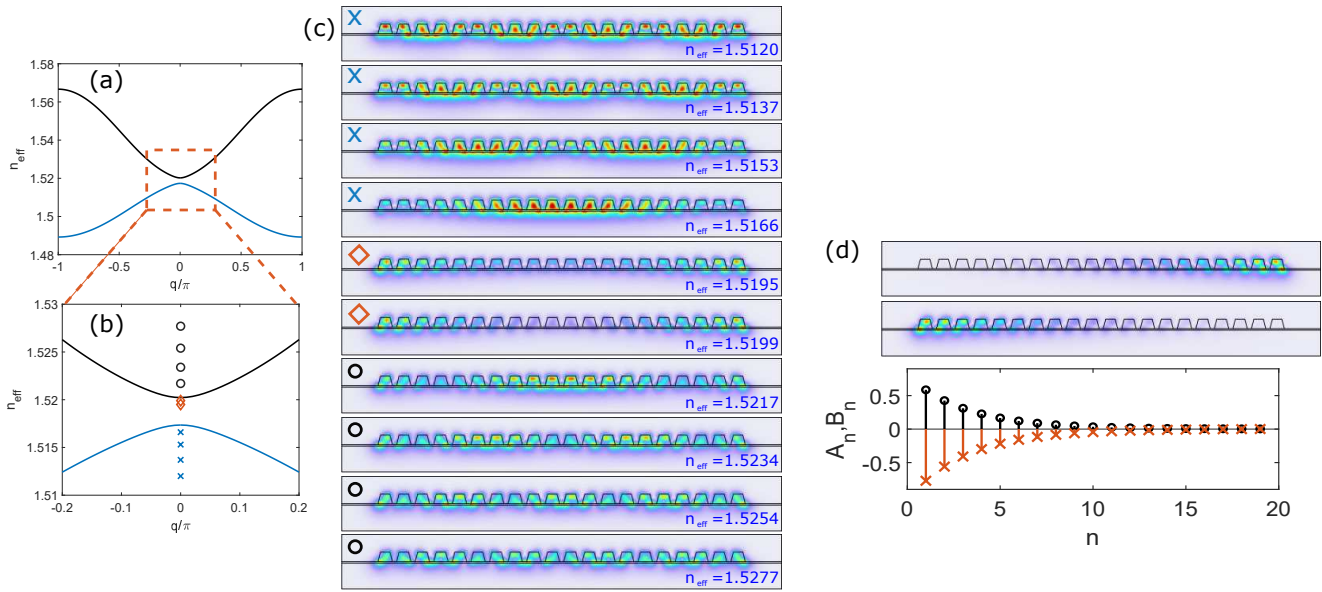}
    \caption{Delocalized and edge TE modes in the geometry as in Fig.~\ref{fig:fig1}(b) at $\lambda=1.5\mu\textrm{m}$, $s=300\textrm{nm}$: (a) the spectrum of the infinite periodic system, and (b) the zoom-in area of the spectrum in a vicinity of the gap; (c) electric field norm of 10 modes of a $N=20$ waveguide array in a vicinity of the gap. The corresponding effective indices are plotted with blue crosses, red diamonds, and black circles in panel (b); modes obtained by adding and subtracting two degenerate Comsol modes indicated by red diamonds in panel (c). The bottom panel displays profile of the corresponding edge mode obtained from the coupled-mode model, Eqs.~(\ref{eq:An},\ref{eq:Bn}).}
    \label{fig:fig5}
\end{figure*}

Similar behaviour is observed with quasi-TE modes  in the second geometry, as in Fig.~\ref{fig:fig1}(b). In Fig.~\ref{fig:fig5}(a) the spectrum of the infinite periodic structure (in a vicinity of the $\textrm{TE}_{01}$ and $\textrm{TE}_{10}$ modes of an isolated waveguide) is shown. This was obtained from COMSOL simulations of the periodic structure. Fig.~\ref{fig:fig5}(b) zooms in the area near the gap. In Fig.~\ref{fig:fig5}(c) we present mode profiles of a finite-size array with $N=20$ waveguides. We picked $10$ modes with indices closest to the gap, the corresponding indices are indicated in Fig.~\ref{fig:fig5}(b) with black circles, blue crosses, and red diamonds. Most of the modes appear to be delocalized, the corresponding indices are within either top (black circles) or bottom (blue crosses) bands of the infinite structure. The two modes indicated by red diamonds fall within the gap. Notably, the two modes have nearly degenerate indices, and their profiles appear to be very similar, with the field intensity being localized at the edges. These are the symmetric and anti-symmetric combinations of the edge modes, as generated by the COMSOL solver due to the degeneracy. The edge modes can thus be reconstructed from these two degenerate modes, as shown in Fig.~\ref{fig:fig5}(c). The bottom panel in Fig.~\ref{fig:fig5}(c) shows the corresponding edge mode as obtained from the coupled-modes model. As before, the two models appear to be in excellent agreement.

We find topological edge modes when the condition 
\begin{equation}
\label{eq:edge_mode_condition}
|n_-|<2|C_-|
\end{equation}
is satisfied.
This inequality compares the mismatch in the effective indices, on the left side, and in the intra-modal coupling coefficients, on the right side, of the two nearly-degenerate modes of an isolated waveguide. Interestingly, the inter-modal coupling coefficient $C_x$ does not enter this condition explicitly, but an interaction between the two families of modes is required to form the edge states. Surprisingly, we discover that for LNOI waveguide arrays the condition in Eq.~(\ref{eq:edge_mode_condition}) is satisfied across large regions of the $(\lambda,s)$ parameter space, leading us to conclude that topological states are a generic feature of this system, see e.g., Fig.~\ref{fig:fig4}(a) and (b). There are two factors which contribute to having large regions of parameter space correspond to topological states. First, we find pairs of nearly degenerate modes (quantified by a small effective index mismatch $n_-$) across large frequency windows due to the combined effects of the material dispersion of Lithium Niobate and the geometric dispersion of the nano-waveguides.  Here we presented two example geometries with different combinations of the first-order TE and TM modes, but we expect this small mismatch to also occur for pairs of other higher-order modes. Second, due to the different parities of the participating modes, the inter-modal coupling coefficients $C_A$ and $C_B$ generally appear to be of opposite signs, thus maximising $2|C_-|=|C_B-C_A|$. As a result, we observe non-trivial topology even when considering pairs of modes with a large detuning $|n_-|$. 

Thus, LNOI waveguide arrays represent a convenient topological photonics platform, in which topology occurs within systems readily fabricated using standard techniques. Significantly, exploiting pairs of near-degenerate modes replaces a two-waveguide unit cell by a single waveguide, thereby decreasing the size of arrays in which topological effects are observed by a factor of two. 
Combined with the strong second-order optical nonlinearity of Lithium Niobate, we find this system especially promising for further studies of nonlinear topological phenomena, such as topological optical parametric oscillations~\cite{Roy2022} or dynamics of two-colour topological edge solitons~\cite{Kartashov2022}.

\bibliography{Topological_photonics}

\end{document}